\documentclass{ifacconf}

\theoremstyle{plain}
\newtheorem{cond}[thm]{Condition}

\usepackage{amssymb}
\usepackage{caption}
\usepackage{subcaption}
\usepackage{multirow}
\usepackage{booktabs}
\usepackage{url}
\usepackage{algorithm}
\usepackage{algpseudocode}
\usepackage{amsfonts}
\usepackage{amsmath}
\usepackage{xcolor}
\usepackage{graphicx}      
\usepackage{natbib}        

\colorlet{BLUE}{blue}
\begin{document}
\begin{frontmatter}

\title{Orthogonal-by-construction augmentation of physics-based input-output models\thanksref{proj}} 

\thanks[proj]{This project has been supported by the Air Force Office of Scientific Research under award number FA8655-23-1-7061. This work is also partly funded by the European Union (ERC, COMPLETE, 101075836). Views and opinions expressed are however those of the authors only and do not necessarily reflect those of the European Union or the European Research Council Executive Agency. Neither the European Union nor the granting authority can be held responsible for them.}

\author[SZTAKI]{Bendegúz M. Györök}
\author[TUe]{Maarten Schoukens}
\author[SZTAKI]{Tamás Péni} 
\author[SZTAKI,TUe]{Roland Tóth}

\address[SZTAKI]{Systems and Control Laboratory, HUN-REN Institute for Computer Science and Control, Budapest, Hungary (e-mails: \{gyorokbende;peni\}@sztaki.hu)}
\address[TUe]{Control Systems Group, Eindhoven University of Technology, Eindhoven, The Netherlands (e-mails: \{m.schoukens;r.toth\}@tue.nl)}

\begin{abstract}                
This paper proposes a novel orthogonal-by-construction parametrization for augmenting physics-based input-output models with a learning component in an additive sense. The parametrization allows to jointly optimize the parameters of the physics-based model and the learning component. Unlike the commonly applied additive (parallel) augmentation structure, the proposed formulation eliminates overlap in representation of the system dynamics, thereby preserving the uniqueness of the estimated physical parameters, ultimately leading to enhanced model interpretability. By theoretical analysis, we show that, under mild conditions, the method is statistically consistent and guarantees recovery of the true physical parameters. With further analysis regarding the asymptotic covariance matrix of the identified parameters, we also prove that the proposed structure provides a clear separation between the physics-based and learning components of the augmentation structure. The effectiveness of the proposed approach is demonstrated through simulation studies, showing accurate reproduction of the data-generating dynamics without sacrificing consistent estimation of the physical parameters.
\end{abstract}


\end{frontmatter}

\section{Introduction}

In recent years, the increasing complexity of engineering systems and the growing performance demands in control applications have intensified the need for accurate nonlinear models. \emph{Discrete-time} (DT) \emph{input-output} (IO) models are one of the most commonly applied structures in system identification as they incorporate a broad spectrum of model classes, ranging from \emph{linear time-invariant} (LTI) \citep{hespanha_linear_2018} and \emph{linear parameter-varying} (LPV) \citep{toth_modeling_2010} models to various nonlinear structures \citep{schoukens_nonlinear_2019}. {\em First-principle} (FP) models can be obtained in DT IO form based on known physical laws and engineering insight. While such physics-based models provide interpretable system descriptions, they often capture only the dominant dynamics, while additional high-complexity effects, such as friction properties or aerodynamic forces, are typically neglected.

As an alternative approach, various data-driven identification methods have been developed for modeling nonlinear systems with an IO model structure \citep{schoukens_nonlinear_2019}. In particular, recent advances employing deep {\em artificial neural networks} (ANNs) have demonstrated superior modeling accuracy compared to conventional approaches \citep{ljung_deep_2020}. However, the practical use of ANN-based black-box models in control-oriented applications, e.g., trajectory planning, remains limited due to their lack of physical interpretability \citep{ljung_perspectives_2010}. Furthermore, ANN-based models typically exhibit poor extrapolation capabilities beyond the range of the training data, and substantial learning effort is often spent on rediscovering system behaviors that are already well understood from first-principles knowledge.

To address these challenges, different strategies have been proposed in the literature, starting from  (light) grey-box modeling \citep{bohlin_practical_2006} till \emph{physics-informed neural networks} (PINNs) \citep{raissi_physics-informed_2019} and \emph{physics-guided neural networks} (PGNNs) \citep{daw_physics-guided_2022}. One of the most promising directions of these hybrid approaches is model augmentation \citep{schon_multi-objective_2022,hoekstra_learning-based_2025}. The augmentation approach aims at combining FP models, i.e., baseline models, with flexible learning components to achieve faster convergence and better model accuracy compared to black-box learning methods \citep{djeumou_neural_2022}. Furthermore, model augmentation produces interpretable models with a clear understanding of how the learning component complements the baseline dynamics.

In this paper, we investigate a widely used model augmentation structure in the literature, namely the additive formulation. This approach connects the physics-based and learning components in parallel and the corresponding parameters of the two components are jointly optimized. However, simultaneously tuning the learning-based and physical parameters results in the two "subcomponents" competing with each other \citep{bolderman_feedforward_2022}. As a result, the ANN can learn relations that could be represented by the FP model, while the baseline parameters can be tuned to physically unrealistic values. This effect undermines physical interpretability of the model estimate and can even compromise the extrapolation capabilities of the resulting model.

This challenge was first addressed in \cite{bolderman_feedforward_2022} in the context of PGNN-based feedforward control by introducing an additional regularization term into the cost function, penalizing deviations of the baseline parameters from their nominal values. This addition to the cost function effectively limits deviations of the baseline parameters compared to their initial values, hence preventing them from reaching a physically unrealistic parameter domain. Despite its simplicity, the approach has provided good experimental results \citep{bolderman_physics-guided_2024}; moreover, the method can be straightforwardly extended for more complex model augmentation structures \citep{hoekstra_learning-based_2025}. Another attractive approach is based on an orthogonal projection-based regularization, introduced in \cite{kon_physics-guided_2022}, also for feedforward control applications. This approach promotes a specific orthogonality between the baseline and learning components via regularization, penalizing when the ANN learns the already known relations represented by the baseline model. This technique has been adapted and generalized for nonlinear system identification in \cite{gyorok_orthogonal_2025}; however, since these approaches promote orthogonality via regularization, inherently, there exists a trade-off between model accuracy and the desired complementarity. Finding the appropriate trade-off parameter (i.e., regularization coefficient) may not be intuitive. To overcome this, we propose a direct parametrization with guaranteed orthogonality between the baseline and learning components on a selected data set without requiring any trade-off parameter. We also show that under certain conditions for the dataset used to impose such an orthogonality property, the tuned baseline parameters converge to their physically true values.

The main contributions of this work are summarized as:
\begin{itemize}
    \item Proposing an orthogonal-by-construction parametrization for additive augmentation of baseline models in input-output form.
    \item Deriving a theoretical error value for the estimated baseline model parameters.
    \item Proving that the proposed model estimator is consistent in the statistical sense and showing that there is zero covariance between the estimated parameters of the baseline and the learning components.
    \item Demonstrating the advantages of the orthogonal model augmentation structure via an extensive identification study.
\end{itemize}

The remainder of the paper is organized as follows: Sect.~\ref{sec:problem_statement} introduces the considered model augmentation problem with the additive structure and model learning setup. Then, Sect.~\ref{sec:orth_param} discusses the problems caused by the non-unique parametrization of the standard additive structure and proposes an orthogonal-by-construction model parametrization that addresses these challenges. In Sect.~\ref{sec:theoretical_anal}, the theoretical analysis of the proposed parametrization is presented. We provide conditions under which the presented model augmentation approach with joint parameter estimation recovers the physically true parameters of the baseline model, followed by the consistency analysis of the method. Sect.~\ref{sec:ident_example} shows two numerical examples, where we demonstrate the effectiveness of the proposed orthogonal-by-construction model augmentation method. Finally, the conclusions on the achieved results are drawn in Sect.~\ref{sec:conclusion}.

\section{Problem statement}\label{sec:problem_statement}

We consider the dynamics of the data-generating system defined by a discrete-time input-output process:
\begin{equation}
\label{eq:DT-IO}
    y_k = f(x_k) + e_k,
\end{equation}
where $k\in\mathbb{Z}$ is the discrete time index, $y_k\in\mathbb{R}^{n_\mathrm{y}}$ is the measured output, $x_k\in\mathbb{R}^{n_\mathrm{x}}$ contains the lagged IO instances with $x_k=\mathrm{vec}(y_{k-n_\mathrm{a}}^{k-1}, u_{k-n_\mathrm{b}}^k)\in\mathbb{R}^{n_\mathrm{x}}$, where $y_{k-n_\mathrm{a}}^{k-1} = \begin{bmatrix}
    y_{k-1}^\top & y_{k-2}^\top & \cdots & y_{k-n_\mathrm{a}}^\top
\end{bmatrix}^\top \in \mathbb{R}^{n_\mathrm{a}n_\mathrm{y}}$ for $n_\mathrm{a}> 0$ and $\emptyset$ otherwise, being the lagged output values, and $u_{k-n_\mathrm{b}}^k$ can be defined similarly for the lagged input values. Moreover, $f:\mathbb{R}^{n_\mathrm{x}}\rightarrow \mathbb{R}^{n_\mathrm{y}}$ is a nonlinear function and $e_k\in\mathbb{R}^{n_\mathrm{y}}$ is represented by a white noise process with finite variance. The formulation of \eqref{eq:DT-IO} represents a wide range of systems depending on $n_\mathrm{a} \geq 0$, $n_\mathrm{b} \geq 0$, called NARX-type systems.

The exact dynamics of \eqref{eq:DT-IO} are not known, but we assume that based on prior knowledge, a physics-based approximative model (baseline model) in a linear-in-the-parameters form is available as
\begin{equation}\label{eq:fp_model}
    \hat{y}_k = \phi(x_k)\theta_\mathrm{b},
\end{equation}
where $\hat{y}_k\in\mathbb{R}^{n_\mathrm{y}}$ is the model output, $\theta_\mathrm{b}\in\mathbb{R}^{n_{\theta_\mathrm{b}}}$ contains the physical parameters, $\phi:\mathbb{R}^{n_\mathrm{x}} \rightarrow \mathbb{R}^{n_\mathrm{y} \times n_\theta}$ is the regressor matrix-function of the baseline model. Furthermore, we assume that based on first-principles modeling, an initial rough estimate for the baseline parameters is available, and is denoted as $\theta_\mathrm{b}^0$.

Since the baseline model only provides an approximation of the dominant dynamics in \eqref{eq:DT-IO}, common practice is to augment it with an additive learning component as
\begin{equation}\label{eqs:additive_struct}
    \hat{y}_k = \phi(x_k)\theta_\mathrm{b} + f_{\theta_\mathrm{a}}^\mathrm{ANN}(x_k),
\end{equation}
where $f^\mathrm{ANN}$ here represents a fully connected, feedforward neural network with $\theta_\mathrm{a}\in\mathbb{R}^{n_{\theta_\mathrm{a}}}$ being the collection of its parameters. Alternatively, $f^\mathrm{ANN}_{\theta_\mathrm{a}}$ can be replaced by any function approximator without loss of generality. Instead of the additive formulation, many other model augmentation structures can be selected from the literature, e.g., see \cite{retzler_learning-based_2024}. However, the additive, i.e., parallel, formulation offers a transparent model structure with clear separation between the baseline and learning components \citep{hoekstra_learning-based_2025}, hence can be an attractive approach for practical applications.

To achieve the best possible data-fit, while simultaneously acquiring as accurate baseline parameters as possible, an efficient approach is to co-estimate $\theta_\mathrm{b}$ and $\theta_\mathrm{a}$ parameters, as proposed in \cite{bolderman_feedforward_2022}. Hence, using a data sequence $\mathcal{D}_N=\left\{(x_i, y_i)\right\}_{i=0}^{N-1}$ generated by \eqref{eq:DT-IO}, the parameters are estimated by minimizing the prediction error loss function expressed as
\begin{equation}\label{eq:cost_fun}
    V_{\mathcal{D}_N}(\theta_\mathrm{b}, \theta_\mathrm{a}) = \frac{1}{N} \|Y - \hat{Y}\|_2^2,
\end{equation}
where $Y=\begin{bmatrix}
    y_0^\top & y_1^\top & \cdots & y_{N-1}^\top
\end{bmatrix}^\top$ are the stacked measured output values. Moreover, $\hat{Y}$ is computed, as
\begin{equation}\label{eq:vectorized_form}
    \underbrace{\begin{bmatrix}
        \hat{y}_0\\ \hat{y}_1\\ \vdots\\ \hat{y}_{N-1}
    \end{bmatrix}}_{\hat{Y}} = \underbrace{\begin{bmatrix}
        \phi(x_0)\\ \phi(x_1)\\ \vdots \\\phi(x_{N-1})
    \end{bmatrix}}_{\Phi} \theta_\mathrm{b} + \underbrace{\begin{bmatrix}
        f_{\theta_\mathrm{a}}^\mathrm{ANN}(x_0)\\ f_{\theta_\mathrm{a}}^\mathrm{ANN}(x_1)\\ \vdots\\ f_{\theta_\mathrm{a}}^\mathrm{ANN}(x_{N-1})
    \end{bmatrix}}_{F^\mathrm{ANN}_{\theta_\mathrm{a}}},
\end{equation}
where $\hat{Y} \in\mathbb{R}^{Nn_\mathrm{y}}$ is the vectorized form of the model responses, $\Phi\in\mathbb{R}^{Nn_\mathrm{y}\times n_{\theta_\mathrm{b}}}$ contains the baseline regressor matrices corresponding to the training data, while $F_{\theta_\mathrm{a}}^\mathrm{ANN}$ contains the learning component terms.

To ensure the feasibility of recovering the physically true parameters of the baseline model, certain conditions must be satisfied. Specifically, we require that $\phi(\cdot)\theta_\mathrm{b}^{(1)}=\phi(\cdot)\theta_\mathrm{b}^{(2)}$ implies $\theta_\mathrm{b}^{(1)}=\theta_\mathrm{b}^{(2)}$, which corresponds to an \emph{identifiably} condition under the functions composing $\phi$ (distinguishability of $\theta$), and we require the input sequence in the data set $\mathcal{D}_N$ to be \emph{weakly persistently exciting} in the sense that
\begin{equation}
\|\Phi(\theta_\mathrm{b}^{(1)} - \theta_\mathrm{b}^{(2)})\|_2^2 = 0
\ \ \Rightarrow \ \
\phi(\cdot)\theta_\mathrm{b}^{(1)} = \phi(\cdot)\theta_\mathrm{b}^{(2)}.
\end{equation}
Under these conditions, the regressor matrix $\Phi$ is full rank. Let $\theta_\mathrm{b}^\ast$ denote the physically true baseline parameters. If $\Phi$ is not full rank, there exists a non-zero vector $p$ such that $\Phi p = 0$. In this case, $\Phi\theta_\mathrm{b}^\ast = \Phi(\theta_\mathrm{b}^\ast+\lambda p)$, where $\lambda\in\mathbb{R}$ is an arbitrary non-zero constant, implying that $\theta_\mathrm{b}^\ast$ is not uniquely identifiably from $\mathcal{D}_N$. 
Therefore, we make the following assumption.
\begin{assum}\label{assum:phi_full_rank}
    The training data set $\mathcal{D}_N$ satisfies
    \begin{equation}
        \mathrm{rank}(\Phi) = n_\theta,
    \end{equation}
    where $\Phi\in\mathbb{R}^{Nn_\mathrm{y}\times n_\theta}$ with $Nn_\mathrm{y}\geq n_\theta$.
\end{assum}
This is a core assumption upon which the subsequent orthogonal parametrization is developed. It should be noted, however, that this condition only guarantees a unique solution of the estimation problem w.r.t. the baseline model. Further discussions on how the learning component influences this property will be provided in Sect.~\ref{sec:orth_param} and \ref{sec:theoretical_anal}.

\section{Orthogonal-by-construction parametrization}\label{sec:orth_param}

\subsection{Non-uniqueness of the parametrization}

Commonly applied function approximators, such as ANNs, employed to parameterize the learning component, are typically overparameterized. As a result, multiple parameter values of $\theta_\mathrm{a}$ can result in the same IO relations of the learning component. This is generally referred to as non-identifiability. More critically, due to the inherent structure of the additive model augmentation in \eqref{eqs:additive_struct}, multiple parameter pairs $(\theta_\mathrm{b}, \theta_\mathrm{a})$ can minimize \eqref{eq:cost_fun} even when the learning component itself is uniquely parameterized. As a consequence, the baseline parameters might be tuned to unrealistic values; hence, the interpretability of the augmented model can be compromised. Example~\ref{exmp:non-uniqueness} demonstrates the effect of this parameter non-uniqueness on the interpretability of the model augmentation structure.

\begin{exmp}\label{exmp:non-uniqueness}
    Consider a data-generating system as $y_k=x_k^\top \theta_\mathrm{b}^\ast$ and an IO baseline model of $\hat{y}_k=x_k^\top\theta_\mathrm{b}$. By using a single linear layer in the ANN, which gives $f^\mathrm{ANN}_{\theta_\mathrm{a}}(x_k) = x_k^\top W$, any $\left(\theta_\mathrm{b},\,W\right)$ pair that satisfies $W + \theta_\mathrm{b} = \theta_\mathrm{b}^\ast$ is a global minimizer of \eqref{eq:cost_fun}.
\end{exmp}

\subsection{Direct parametrization of orthogonal subcomponents}
The illustrated overparametrization problem means that the learning component can identify such relations that otherwise could be captured by the baseline model. This naturally conflicts with the aim of model augmentation, namely to incorporate as much physics-based information into the (interpretable) baseline model as possible. With non-unique $\theta_\mathrm{b}$, the baseline model could lose its physical meaning, and might even compromise the extrapolation capabilities of the final model. To address this challenge, first, we introduce the following parameter:
\begin{align}\label{eq:theta_aux}
    \theta_\mathrm{aux} = \left(\Phi^\top \Phi\right)^{-1}\Phi^\top F_{\theta_\mathrm{a}}^\mathrm{ANN}.
\end{align}
The specified data informativity condition in Sect.~\ref{sec:problem_statement} implies that $\Phi$ is full rank; hence, the inverse $\left(\Phi^\top \Phi\right)^{-1}$ exists. In fact, $\left(\Phi^\top \Phi\right)^{-1}\Phi$ corresponds to the Moore-Penrose pseudo inverse of $\Phi$. With the introduced parameter $\theta_\mathrm{aux}$, the prediction map \eqref{eq:vectorized_form} is modified as
\begin{equation}\label{eq:orth_param}
    \hat{Y} = \Phi \theta_\mathrm{b} + \underbrace{F_{\theta_\mathrm{a}}^\mathrm{ANN} - \Phi \theta_\mathrm{aux}}_{\tilde{F}^\mathrm{ANN}_{\theta_\mathrm{a}}},
\end{equation}
where $\tilde{F}^\mathrm{ANN}_{\theta_\mathrm{a}}$ denotes the vectorized formulation of the projected learning component. Substituting $\theta_\mathrm{aux}$ based on \eqref{eq:theta_aux} into \eqref{eq:orth_param}, the proposed parametrization for the learning component can be expressed as
\begin{equation}\label{eq:orth_full_param}
    \tilde{F}_{\theta_\mathrm{a}}^\mathrm{ANN} = \left[I - \Phi\left(\Phi^\top\Phi\right)^{-1}\Phi^\top\right] F^\mathrm{ANN}_{\theta_\mathrm{a}}.
\end{equation}

The presented model structure ensures guaranteed orthogonality between the baseline and learning components over the training data, as shown in Lemma~\ref{property:orthogonality}.

\begin{lem}\label{property:orthogonality}
    Following the parametrization outlined in \eqref{eq:orth_param}, orthogonality between the baseline and projected learning component is guaranteed on the training set, i.e.,
    \begin{equation}\label{eq:orth_property}
        \Phi^\top \tilde{F}_{\theta_\mathrm{a}}^\mathrm{ANN} = 0.
    \end{equation}
\end{lem}
\begin{pf}
    Substituting \eqref{eq:orth_full_param} into \eqref{eq:orth_property}, we arrive to
    \begin{equation}
        \Phi^\top F_{\theta_\mathrm{a}}^\mathrm{ANN} - \underbrace{\Phi^\top\Phi \left(\Phi^\top \Phi\right)^{-1}}_{I}\Phi^\top F_{\theta_\mathrm{a}}^\mathrm{ANN} = 0.\vspace{-12pt}
    \end{equation} \hfill $\blacksquare$
\end{pf}

Alternatively, $\theta_\mathrm{aux}$ in \eqref{eq:theta_aux} can be constructed by using any auxiliary evaluation of the regressor  $\phi$ that can either be on a synthetically generated data set or a subset of the estimation data. In the remainder of this paper, we will assume that the whole training data set is utilized when constructing \eqref{eq:theta_aux}, but keep in mind that the proposed methodology is not restricted to this scenario.

For the applied parametrization, model training now results in the estimated $\hat{\theta}_\mathrm{b}$, $\hat{\theta}_\mathrm{a}$ parameters and, moreover, a fixed $\hat{\theta}_\mathrm{aux}$ value. This is due to the applied orthogonal projection depending on the data distribution of $\mathcal{D}_N$. Hence, after training, prediction on new test data can be computed as
\begin{equation}
    \hat{y}_k = \phi(x_k)\hat{\theta}_\mathrm{b} + f^\mathrm{ANN}_{\hat{\theta}_\mathrm{a}}(x_k) - \phi(x_k)\hat{\theta}_\mathrm{aux}. 
\end{equation}

\section{Theoretical analysis}\label{sec:theoretical_anal}

\subsection{Recovery of the baseline parameters}\label{sec:baseline_recovery}

To analyze theoretically the error of the estimated baseline parameters, first, we reformulate the data-generating system \eqref{eq:DT-IO} as
\begin{equation}\label{eq:data_gen_sy_reformed}
    y_k = \phi(x_k)\theta_\mathrm{b}^\ast + \delta(x_k) + e_k,
\end{equation}
where $\delta : \mathbb{R}^{n_\mathrm{x}} \rightarrow \mathbb{R}^{n_\mathrm{y}}$ represents the unmodeled terms, while $\theta_\mathrm{b}^\ast\in\mathbb{R}^{n_\theta}$ denotes the physically true baseline parameters. Then, a similar vectorized form of the true dynamics on the training data can be provided as in \eqref{eq:vectorized_form}:
\begin{equation}\label{eq:data_gen_sys_reformed_vect}
    Y = \Phi\theta_\mathrm{b}^\ast + \Delta + E,
\end{equation}
where $Y\in\mathbb{R}^{Nn_\mathrm{y}}$ is the vectorized form of the system outputs, and $\Delta\in\mathbb{R}^{Nn_\mathrm{y}}$, $E\in\mathbb{R}^{Nn_\mathrm{y}}$ are the collection of the $\delta(x_k)$ and $e_k$ terms, respectively.

Achieving orthogonal subcomponents is only realistic if the baseline model regressor matrix function is in fact orthogonal to the unmodeled terms on a task-specific operating domain $x_k\in\mathbb{X}\subseteq \mathbb{R}^{n_\mathrm{x}}$. Thus,
\begin{equation}\label{eq:phi-delta-orthog-functions}
    \int_{x\in\mathbb{X}} \phi^\top(x) \delta(x)~\mathrm{d}x = 0.
\end{equation}

Moreover, we require that the above-defined orthogonality is reflected in the gathered data.

\begin{cond}\label{cond:special_PE}
The data set $\mathcal{D}_N$ generated by \eqref{eq:data_gen_sy_reformed} satisfies
    \begin{equation}\label{eq:phi-delta-discrete-orthogonality}
    \sum_{i=0}^{N-1} \phi^\top(x_i)\delta(x_i) = 0.
\end{equation}
\end{cond}
For certain basis functions in $\phi$ and $\delta$ (e.g., orthogonal polynomials), \eqref{eq:phi-delta-orthog-functions} implies that with $N\to\infty$ there always exists an appropriate selection of regressor points $x_k$ to satisfy Condition~\ref{cond:special_PE}. For other scenarios, a dedicated experiment design is necessary. An alternative interpretation is that \eqref{eq:phi-delta-orthog-functions} is the identifiability criterion of model class \eqref{eq:orth_param}, while \eqref{eq:phi-delta-discrete-orthogonality} is a specific excitation condition for the considered identification problem.
\begin{rem}
    If the structure of $\delta$ can be inferred from prior physical knowledge, it is possible to specifically design an input sequence that satisfies Condition~\ref{cond:special_PE} (at least with respect to the suspected unmodeled terms). This is a common scenario in practical applications. For example, in aerial robotics, the drag force is known to depend quadratically on the relative velocity of the object \citep{huang_leveraging_2024}. However, the associated aerodynamic computations are often complex, and this force term is typically left unmodeled. Incorporating these physics-based insights can, however, guide the design of experiments that fulfil Condition~\ref{cond:special_PE}. 
\end{rem}
Next, we assume that the minimization of \eqref{eq:cost_fun} results in such $\theta_\mathrm{b}$, $\theta_\mathrm{a}$ estimates for which the relations of \eqref{eq:data_gen_sy_reformed} are exactly recovered on the training data.

\begin{assum}\label{assum:data_gen_sys_recovered}
    The identified model recovers the dynamics of the data-generating system \eqref{eq:data_gen_sy_reformed} on the training data set $\mathcal{D}_N$, as
    \begin{equation}\label{eq:data_gen_sy_recovered}
        \Phi \theta_\mathrm{b}^\ast + \Delta = \Phi \hat{\theta}_\mathrm{b} + \tilde{F}^\mathrm{ANN}_{\hat{\theta}_\mathrm{a}},
    \end{equation}
    where $\hat{\theta}_\mathrm{b}$ and $\hat{\theta}_\mathrm{a}$ are the estimated parameters of the baseline and learning component, respectively.
\end{assum}
Note that Assumption~\ref{assum:data_gen_sys_recovered} is only realistic for a finite $N$ number of data points if the training data does not contain any noise. Later, in Sect.~\ref{subsec:consistency}, we will show that under certain conditions, Assumption~\ref{assum:data_gen_sys_recovered} holds as $N\to\infty$ even when noise is present in the data, i.e., we will prove consistency of the estimator. Before moving to the noisy scenario, we show that the proposed orthogonal parametrization recovers the physically true parameters of the baseline model if the training data is noiseless.
\begin{thm}\label{thm:error_val}
    Under Assumptions~\ref{assum:phi_full_rank} and \ref{assum:data_gen_sys_recovered}, if Condition~\ref{cond:special_PE} is satisfied, then the estimation error of the baseline parameters is zero, i.e., $\hat{\theta}_\mathrm{b}=\theta_\mathrm{b}^\ast$.
\end{thm}
\begin{pf}
    Assumption~\ref{assum:data_gen_sys_recovered} implies that the relations of the data-generating system are exactly recovered on the training data. Assumption~\ref{assum:phi_full_rank} dictates that parametrization \eqref{eq:orth_param} exists and the true baseline parameters can be identified based on $\mathcal{D}_N$. Hence, substituting \eqref{eq:orth_param} to \eqref{eq:data_gen_sy_recovered} leads to
    \begin{equation}
        \Phi \theta_\mathrm{b}^\ast + \Delta = \Phi\hat{\theta}_\mathrm{b} + \left[ I - \Phi \left(\Phi^\top \Phi\right)^{-1}\Phi^\top\right]F_{\hat{\theta}\mathrm{a}}^\mathrm{ANN}.
    \end{equation}
    Left multiplying both sides with $\Phi^\top$, then dropping out terms similarly as in the proof of Lemma~\ref{property:orthogonality}, we arrive to
    \begin{equation}
        \Phi^\top \Phi \theta_\mathrm{b}^\ast + \Phi^\top \Delta = \Phi^\top\Phi\hat{\theta}_\mathrm{b}.
    \end{equation}
    Re-arranging the terms and taking the $\ell_2$ norm of both sides leads to
        \begin{equation}\label{eq:error_val}
        \|\theta_\mathrm{b}^\ast - \hat{\theta}_\mathrm{b}\|_2 = \|\left(\Phi^\top \Phi\right)^{-1} \Phi^\top \Delta\|_2,
    \end{equation}
    which is exactly zero, since $\Phi^\top \Delta=0$ according to Condition~\ref{cond:special_PE}.\hfill $\blacksquare$
\end{pf}

\subsection{Comparison with the standard parametrization}

Two important factors can be derived from Theorem~\ref{thm:error_val}. First, when Assumptions \ref{assum:phi_full_rank} and \ref{assum:data_gen_sys_recovered} hold, but Condition~\ref{cond:special_PE} does not, the true baseline parameters can not be recovered, but the acquired $\hat{\theta}_\mathrm{b}$ value is unique and the estimation error is given by \eqref{eq:error_val}. Secondly, using the standard additive structure \eqref{eqs:additive_struct} can result in larger baseline parameter errors (in a worst-case sense) than the orthogonal parametrization, even when Condition \ref{cond:special_PE} is not satisfied. To show this, we make the following reasonable assumption.

\begin{assum}\label{assum:baseline_mdl_dominant}
    The baseline model contains the dominant characteristics of the data-generating system; 
    hence, for the given
    data sequence in $\mathcal{D}_N$, it holds that
    \begin{equation}
        \|\Phi\theta_\mathrm{b}^\ast \|_2 > \|\Delta\|_2.
    \end{equation}
\end{assum}

Let the error of the baseline parameters with the orthogonal-by-construction parametrization be denoted by $e_{\theta_\mathrm{b}}^\mathrm{orth}$, which is given by \eqref{eq:error_val}, based on Theorem~\ref{thm:error_val}. Similarly, let us introduce a notation for the same error value, but without the proposed parametrization. However, without orthogonalization, $\hat{\theta}_\mathrm{b}$ is non-unique, hence, when Assumption~\ref{assum:data_gen_sys_recovered} holds, the following equivalence set can be defined:
\begin{multline}\label{eq:eqivalence_set_std}
    \Theta_\mathrm{b} \times \Theta_\mathrm{a}= \{(\theta_\mathrm{b}, \theta_\mathrm{a}) \mid \phi(x_k)\theta_\mathrm{b}+f^\mathrm{ANN}_{\theta_\mathrm{a}}(x_k) = \\\phi(x_k)\theta_\mathrm{b}^\ast + \delta(x_k),\quad \forall x_k\in\mathcal{D}_N\}.
\end{multline}
Then, $e_{\theta_\mathrm{b}}^\mathrm{std}$ denotes the upper bound of the error value, as
\begin{equation}
    e_{\theta_\mathrm{b}}^\mathrm{std} = \sup_{\hat{\theta}_\mathrm{b}\in\Theta_\mathrm{b}} \|\theta_\mathrm{b}^\ast - \hat{\theta}_\mathrm{b}\|_2.
\end{equation}

Now we can use a similar analysis as in Theorem~\ref{thm:error_val}, however, the learning component does not drop out when left-multiplying the expression with $\Phi^\top$, and the following error characterization holds under Assumption~\ref{assum:data_gen_sys_recovered}:
\begin{equation}\label{eq:baseline_error_simple_augm}
    \|\theta_\mathrm{b}^\ast - \hat{\theta}_\mathrm{b}\|_2 = \|\left(\Phi^\top \Phi\right)^{-1} \Phi^\top \left(F_{\theta_\mathrm{a}}^\mathrm{ANN} - \Delta\right)\|_2.
\end{equation}
Then, it is enough to show that there exists at least one parametrization for the standard model augmentation approach that satisfies Assumption~\ref{assum:data_gen_sys_recovered}, but provides larger baseline recovery error than the orthogonal-by-construction method. Consider the case when the baseline component is estimated to be zero, $\theta_\mathrm{b} = 0$. Then, based on Assumption~\ref{assum:data_gen_sys_recovered} and the defined equivalence set in \eqref{eq:eqivalence_set_std}, the learning component becomes $F_{\theta_\mathrm{a}}^\mathrm{ANN}=\Phi\theta_\mathrm{b}^\ast + \Delta$. Using this specific scenario and Assumption~\ref{assum:baseline_mdl_dominant}, comparing the formulation of \eqref{eq:baseline_error_simple_augm} with \eqref{eq:error_val},
it straightforwardly follows that $e_{\theta_\mathrm{b}}^\mathrm{base} > e_{\theta_\mathrm{b}}^\mathrm{orth}$.

\begin{rem}
    The error of the estimated baseline parameters in \eqref{eq:baseline_error_simple_augm} is similar to the one in \cite{donati_combining_2025}, which also follows intuitively. When the dynamics of \eqref{eq:data_gen_sy_reformed} are recovered on the training data, the exact baseline parameters are found only if the learning component identifies the unmodeled terms and nothing else. If the ANN learns parts that the baseline model could represent, the true baseline parameters can not be retained.
\end{rem}

\subsection{Consistency analysis}\label{subsec:consistency}

In the previous derivations, we assumed that the model training results in such parameters for which the exact relations of the data-generating system are recovered on the training set. Now, we will provide certain conditions under which this assumption is satisfied for $N\to\infty$, i.e., we will show consistency of the estimator following the arguments of \cite{ljung_convergence_1978}. For that, first we require that the data-generating system \eqref{eq:DT-IO} is stable.

\begin{cond}[Stable data-generating system]\label{cond:stable_sys}
    The data-generating system \eqref{eq:DT-IO} has the property that, for any $\rho>0$, there exist a $C(\rho)\in\left[0, \infty\right)$, and a $\lambda\in\left[0,1\right)$, such that
    \begin{equation}
        \mathbb{E}_e \left\{\|y_k - \tilde{y}_k\|_2^4\right\} < C(\rho) \lambda^{k-k_\mathrm{o}}, \quad \forall k \geq k_\mathrm{o}
    \end{equation}
    under any $k_\mathrm{o} \geq 0$, $x_0, \tilde{x}_0 \in \mathbb{R}^{n_\mathrm{x}}$ with $\|x_0 - \tilde{x}_0\|_2 < \rho$, and $\{(u_i, e_i)\}_{i=0}^\infty\in\mathcal{W}_{[0, \infty]}$, where $\mathcal{W}_{[0, \infty]}$ denotes the $\sigma$-algebra associated with the random variables $\{(u_i, e_i)\}_{i=0}^\infty$; moreover, the random variables $y_k$ and $\tilde{y}_k$ satisfy \eqref{eq:DT-IO} with the same $(u_k, e_k)$, but with $x_{k_\mathrm{o}}=x_0$ and $\tilde{x}_{k_\mathrm{o}}=\tilde{x}_0$.
\end{cond}

Next, we make assumptions on the representation capability of the applied model parametrization. Let us denote the model structure represented by \eqref{eq:orth_param} as $M_\theta$ with $\theta=\mathrm{vec}(\theta_\mathrm{b}, \theta_\mathrm{a})\in\mathbb{R}^{n_\theta}$. Furthermore, we assume that $\theta$ is restricted to vary in a compact set $\Theta\subset \mathbb{R}^{n_\theta}$, hence, the considered model set is given by $\mathcal{M}=\{M_\theta \mid \theta\in\Theta\}$. For a given model structure $M_\theta$, the corresponding \emph{1-step-ahead predictor} can be expressed according to \eqref{eq:orth_param}, as
\begin{equation}\label{eq:OSA_pred}
    \hat{y}_k^\mathrm{pred} = \gamma_k(\theta, \{y_i\}_{i=-n_\mathrm{a}}^{k-1}, \{u_i\}_{i=-n_\mathrm{b}}^{k}).
\end{equation}
We take a further assumption that $\gamma_k$ is differentiable w.r.t. $\theta$ everywhere on an open neighborhood $\breve{\Theta}$ of $\Theta$. In practice, only such parametrizations are considered for which automatic differentiation is available; hence, this is only a technical condition. Moreover, we require $\gamma_k$ to be stable w.r.t. perturbations regarding the data set, to guarantee convergence of the predictor.

\begin{cond}[Stable predictor]\label{cond:pred_stab}
    There exist a $C\in[0,\infty)$ and a $\lambda\in[0,1)$ such that, for any $k\geq 0$ and $\theta\in\breve{\Theta}$, and any $\{u_i, y_i\}_{i=-n}^k$, $\{\tilde{u}_i, \tilde{y}_i\}_{i=-n}^k$ with $n=\mathrm{max}(n_\mathrm{a}, n_\mathrm{b}$), the predictor $\gamma_k$ satisfies that
    \begin{multline}\label{eq:pred_stab}
        \|\gamma_k(\theta, \{y_i\}_{i=-n_\mathrm{a}}^{k-1}, \{u_i\}_{i=-n_\mathrm{b}}^{k}) - \\\gamma_k(\theta, \{\tilde{y}_i\}_{i=-n_\mathrm{a}}^{k-1}, \{\tilde{u}_i\}_{i=-n_\mathrm{b}}^{k})\|_2 \leq C \Gamma_k,
    \end{multline}
    where $\Gamma_k = \sum_{i=-n}^k \lambda^{k-i} (\|u_i-\tilde{u}_i\|_2 + \|y_i-\tilde{y}_i\|_2)$; moreover, $\|\gamma(\theta, \{0\}_{i=-n_\mathrm{a}}^{k-1}, \{0\}_{i=-n_\mathrm{b}}^{k})\|_2 \leq C$. Furthermore, \eqref{eq:pred_stab} is also satisfied by $\frac{\partial}{\partial\theta}\gamma_k$.
\end{cond}

\begin{thm}[Convergence]\label{thm:convergence}
    Consider the data-generating system \eqref{eq:DT-IO} satisfying Condition~\ref{cond:stable_sys} with a quasi-stationary $u$ independent of the white noise process $e$. Given the model set $\mathcal{M}$ defined by \eqref{eq:orth_param} satisfies Condition~\ref{cond:pred_stab}, then
    \begin{equation}
        \sup_{\mathrm{vec}(\theta_\mathrm{b}, \theta_\mathrm{a})\in\Theta} \|V_{\mathcal{D}_N}(\theta_\mathrm{b}, \theta_\mathrm{a}) - \bar{V}(\theta_\mathrm{b}, \theta_\mathrm{a})\|_2 \to 0,
    \end{equation}
    with probability 1 as $N\to\infty$, where $\bar{V}(\theta_\mathrm{b}, \theta_\mathrm{a}) = \lim_{N\to\infty} \frac{1}{N} \mathbb{E} \{\|Y-\hat{Y}\|_2^2$\}.
\end{thm}
\begin{pf}
    The identification criterion given by \eqref{eq:cost_fun} satisfies Condition C1 in \cite{ljung_convergence_1978}, hence the proof of \cite[Lemma 3.1]{ljung_convergence_1978} applies for the considered case.\hfill $\blacksquare$
\end{pf}
\begin{rem}
    In practice, Condition~\ref{cond:pred_stab} serves only as a technical condition. After acquiring the estimate of the model parameters $\hat{\theta}$, it can be numerically checked whether the given model realization satisfies this requirement. For a strict guarantee of satisfying Condition~\ref{cond:pred_stab}, a so-called stable-by-design parametrization can be applied. See, e.g., \cite{revay_recurrent_2024,kon_unconstrained_2025}. Extending these methods for the model augmentation setting is an open question and can be the topic of future research.
\end{rem}

Similarly, as in Sect.~\ref{sec:baseline_recovery}, we assume that system \eqref{eq:data_gen_sy_reformed} belongs to model class $\mathcal{M}$. Due to the overparametrization of the learning problem, we define an equivalence set $\Theta^\ast \subset \Theta$ as in \eqref{eq:eqivalence_set_std}, which contains all $\theta^\ast\in\Theta$ for which $M_{\theta_\ast}$ is equivalent to the data-generating system. Later, we will verify that the baseline part for all $\theta^\ast\in\Theta^\ast$ remains unique by applying the orthogonal-by-construction parameterization. In order to show that, we require that non-equivalent models can be distinguished in $\mathcal{M}$ based on $\mathcal{D}_N$.

\begin{cond}[Persistency of excitation]\label{cond:PE}
    Given model set $\mathcal{M}=\{M_\theta\mid \theta\in\Theta\}$, we call the input sequence $\{u_i\}_{i=0}^{N-1}$ in $\mathcal{D}_N$ weakly persistently exciting, if for all parametrizations given by $\theta_1, \theta_2\in\Theta$ for which the function mapping is unequal, i.e., $V_{(\cdot)}(\theta_1) \neq V_{(\cdot)}(\theta_2)$, we have
    \begin{equation}
        V_{\mathcal{D}_N}(\theta_1) \neq V_{\mathcal{D}_N}(\theta_2),
    \end{equation}
    with probability 1.
\end{cond}

For linear time-invariant models, the classical \emph{persistency of excitation} (PE) condition that is characterized solely on the input trajectory in the open loop case, implies a fundamental informativity requirement, i.e., distinguishability of (equivalence sets of) parameters under the data \citep{Gevers2009}. For NARX-type models with linear-in-the-parameters form as \eqref{eq:fp_model}, the PE condition requires the regressor matrix composed of nonlinear basis functions to be of full column rank. Unlike for linear systems, this condition depends not only on the input signal, but also on the resulting output trajectories. Consequently, PE requirements must be assessed at the level of the nonlinear regressor. Considering the full nonlinear model augmentation structure of \eqref{eqs:additive_struct} further complicates the PE condition. In practice, space-filling and model-based excitation designs (see, e.g., \citep{yuhan_space-filling_2025}) can be applied to select inputs for maximizing data information content over the regressor space.

Lastly, to prove consistency, we need to show that any element of $\Theta_\ast$ has minimal cost with $N\to\infty$.

\begin{lem}[Minimal cost]\label{lem:minimal_cost}
    If $\Theta_\ast \neq \varnothing$, then the minimum of the limit $\lim_{N\to\infty} V_{\mathcal{D}_N}(\theta_\mathrm{b}, \theta_\mathrm{a})$ is reached only when $\mathrm{vec}(\theta_\mathrm{b}, \theta_\mathrm{a}) \in \Theta_\ast$.
\end{lem}
\begin{pf}
    Substituting \eqref{eq:data_gen_sys_reformed_vect} and \eqref{eq:orth_param} into the cost function \eqref{eq:cost_fun} gives
    \begin{equation}
        \frac{1}{N}\|\underbrace{\Phi\theta_\mathrm{b}^\ast + \Delta - \Phi\theta_\mathrm{b} - \tilde{F}_{\theta_\mathrm{a}}^\mathrm{ANN}}_{\varepsilon} + E\|_2^2,
    \end{equation}
    which can be reformulated as
    \begin{equation}\label{eq:cost_fun_N_infty}
        \frac{1}{N}\|\varepsilon\|_2^2 + \frac{2}{N}\varepsilon^\top E \varepsilon + \frac{1}{N}\|E\|_2^2.
    \end{equation}
    As $N\to\infty$, the sample distribution of $\{e_k\}_{k=0}^{N-1}$ will converge to the original white noise distribution of $e_k$ with (finite) variance $\Sigma_\mathrm{e}$. Thus, the second term in \eqref{eq:cost_fun_N_infty} is equal to zero, since $e_k$ is uncorrelated with $(\phi(x_k)\theta_\mathrm{b}^\ast + \delta(x_k)-\hat{y}_k)$. Moreover, $N\to\infty$ also implies that $\frac{1}{N}\|E\|_2^2\to\mathrm{trace}(\Sigma_\mathrm{e})$. Since the first term in \eqref{eq:cost_fun_N_infty} is non-negative,
    \begin{equation}
        \lim_{N\to\infty} \frac{1}{N} \|Y-\hat{Y}\|_2^2 \geq \mathrm{trace}(\Sigma_\mathrm{e}),
    \end{equation}
    where equality (i.e., the minimal cost of the identification criterion) holds when $\varepsilon=0$, thus, when the identified model recovers the relations of the data-generating system, i.e., $\mathrm{vec}(\theta_\mathrm{b}, \theta_\mathrm{a})\in\Theta_\ast$.\hfill $\blacksquare$
\end{pf}

\begin{thm}[Consistency]\label{thm:consistency}
    Under the conditions of Theorem~\ref{thm:convergence} and Lemma~\ref{lem:minimal_cost} and Conditions~\ref{cond:special_PE} and \ref{cond:PE},
    \begin{align}
        \lim_{N\to\infty} \hat{\theta}^N &\in \Theta^\ast,\label{eq:consistency}\\
        \lim_{N\to\infty} \hat{\theta}_\mathrm{b}^N &= \theta_\mathrm{b}^\ast,\label{eq:baseline_consistency}
    \end{align}
    with probability 1, where
    \begin{equation}
        \hat{\theta}^N = \mathrm{vec}(\hat{\theta}_\mathrm{b}^N, \hat{\theta}_\mathrm{a}^N) = \arg\min_{\mathrm{vec}(\theta_\mathrm{b}, \theta_\mathrm{a})\in\Theta} V_{\mathcal{D}_N}(\theta_\mathrm{b}, \theta_\mathrm{a}).
    \end{equation}
\end{thm}
\begin{pf}
    For the proof of \eqref{eq:consistency}, see Lemma 4.1 in \cite{ljung_convergence_1978}. Note that the applied loss function \eqref{eq:cost_fun} fulfills Condition (4.4) in \cite{ljung_convergence_1978}. To prove \eqref{eq:baseline_consistency}, refer back to Theorem~\ref{thm:error_val}. With the conditions of Lemma~\ref{lem:minimal_cost}, Assumption~\ref{assum:data_gen_sys_recovered} is trivially satisfied; hence, \eqref{eq:baseline_consistency} holds under Condition~\ref{cond:special_PE}.\hfill $\blacksquare$
\end{pf}

\begin{rem}
    For the noiseless case, i.e., when $e_k\equiv 0$, the error expression for $\theta_\mathrm{b}$ in \eqref{eq:error_val} holds for all $N$ provided that all other conditions of Theorem~\ref{thm:consistency} are satisfied. In this scenario, attaining the global minimum of the cost function implies that the exact dynamics of the data-generating system are recovered on the training data. In contrast, when $e_k \neq 0$, this equivalence holds only in a statistical sense, which motivates the asymptotic ($N\to\infty$) nature of \eqref{eq:baseline_consistency}.
\end{rem}

\subsection{Covariance of the model parameters}

Finally, we show that, due to the orthogonality between the baseline and learning components, the proposed parametrization results in zero covariance between the estimated $\theta_\mathrm{b}$ and $\theta_\mathrm{a}$ parameters. Under the conditions of the consistency results, the asymptotic distribution of $\hat{\theta}^N$ can be expressed 
w.r.t. to $\lim_{N\rightarrow\infty}\hat{\theta}^N=\theta^\ast\in \Theta^\ast$ as $\sqrt{N}(\hat{\theta}^N-\theta^\ast) \in \mathrm{As}\ \mathcal{N}(0,P_\theta)$ \citep{ljung_system_1998}. Then, under the considered quadratic loss function, 
\begin{multline}
    P_\theta = \left[\bar{\mathbb{E}}\{\psi^\top_k(\theta_\ast)\psi_k(\theta_\ast)\}\right]^{-1} \left[\bar{\mathbb{E}}\{2\psi^\top_k(\theta_\ast)\Sigma_0\psi_k(\theta_\ast)\}\right] \\\left[\bar{\mathbb{E}}\{\psi^\top_k(\theta_\ast)\psi_k(\theta_\ast)\}\right]^{-1},
\end{multline}
where $\psi_k(\theta_\ast) \in\mathbb{R}^{n_\mathrm{y}\times n_\theta}$ is the Jacobian matrix\footnote{Note that $\psi_k$ is defined with a different dimensional notation compared to \cite{ljung_system_1998} for practical reasons.} $\partial\hat{y}_k/\partial\theta$. Based on $\mathcal{D}_N$ and parameter estimate $\hat{\theta}^N$, $P_\theta$ can be estimated as
\begin{multline}\label{eq:cov_approx}
    \hat{P}^N = \left[\frac{1}{N}\sum_{k=0}^{N-1}\psi_k^\top(\hat{\theta}^N)\psi_k(\hat{\theta}^N)\right]^{-1} \\\left[\frac{2}{N}\sum_{k=0}^{N-1}\psi^\top_k(\hat{\theta}^N)\hat{\Sigma}^N\psi_k(\hat{\theta}^N)\right] \left[\frac{1}{N}\sum_{k=0}^{N-1}\psi^\top_k(\hat{\theta}^N)\psi_k(\hat{\theta}^N)\right]^{-1},
\end{multline}
where $\hat{\Sigma}^N=(1/N) \sum_{k=0}^{N-1}(y_k-\hat{y}_k)(y_k-\hat{y}_k)^\top$. This gives an approximation for the covariance of the parameters, as $\mathrm{Cov}(\hat{\theta}^N)\simeq \frac{1}{N} \hat{P}^N$.

\begin{thm}[Zero covariance]\label{thm:zero_cov}
    Under the conditions of Theorem~\ref{thm:consistency} and $\Sigma_\mathrm{e}$ being diagonal, the covariance between the estimated baseline and learning component parameters, $\hat{\theta}_\mathrm{b}^N$ and $\hat{\theta}_\mathrm{a}^N$, is zero.
\end{thm}
\begin{pf}
According to Theorem~\ref{thm:consistency} the approximation in \eqref{eq:cov_approx} is valid. Then, as the parameter vector $\theta$ is separated into the baseline and learning parts, the gradient $\partial\hat{y}_k/\partial\theta$ can be computed separately:
\begin{equation}
    \frac{\partial \hat{y}_k}{\partial \theta_\mathrm{b}} = \phi(x_k),\quad
    \frac{\partial \hat{y}_k}{\partial \theta_\mathrm{a}} = J_f(x_k),
\end{equation}
where $J_f$ denotes the Jacobian of $\tilde{f}^\mathrm{ANN}_{\theta_\mathrm{a}}(x_k)$ w.r.t $\theta_\mathrm{a}$, evaluated at $\hat{\theta}^N$. Since $\psi_k$ is computed on the training data set, $\tilde{f}^\mathrm{ANN}_{\theta_\mathrm{a}}$ is orthogonal to the subspace spanned by $\Phi$; moreover, all changes in the projected ANN output remain orthogonal to the columns of $\Phi$, hence $\phi^\top J_f = 0$. Then, for all data points in $\mathcal{D}_N$
\begin{equation}\label{eq:Hessian}
\psi_k^\top(\hat{\theta}^N)\psi_k(\hat{\theta}^N) = \begin{bmatrix}
    \phi^\top(x_k)\phi(x_k) & 0\\
    0 & J_f^\top(x_k) J_f(x_k)
\end{bmatrix} .
\end{equation}
The expression in \eqref{eq:Hessian} is block-diagonal, hence $\sum_{k=0}^N\psi_k^\top\psi_k$ is a sum of block-diagonal matrices with the same structure. Assuming both blocks are full rank, the first and third terms in \eqref{eq:cov_approx} can be computed by separately inverting the two blocks, again resulting in the same block-diagonal structure. With the noise process $e_k$ affecting each output channel being uncorrelated, it is reasonable to assume that $\hat{\Sigma}^N$ in \eqref{eq:cov_approx} is diagonal. Thus, the middle term in \eqref{eq:cov_approx} also follows the same block diagonal structure, ultimately causing $\mathrm{Cov(\hat{\theta}^N)}$ being block-diagonal, hence the zero covariance between $\hat{\theta}_\mathrm{b}^N$ and $\hat{\theta}_\mathrm{a}^N$.\hfill $\blacksquare$
\end{pf}

Zero covariance between $\hat{\theta}_\mathrm{b}$ and $\hat{\theta}_\mathrm{a}$ implies that the applied parameter initialization of the ANN weights does not affect the estimation of the baseline parameters, maintaining a clear separation between the two submodels and promoting interpretable model augmentation.

\section[Identification examples]{Identification examples}\label{sec:ident_example}
This section presents two identification examples\footnote{The used data and code for both examples are available at: \url{https://github.com/AIMotionLab-SZTAKI/orthogonal-IO-augm}} that demonstrate the effectiveness of the proposed orthogonal-by-construction model augmentation structure. The first example considers a static nonlinear data-generating system and illustrates the theoretical properties derived in previous sections. The second example involves a physics-motivated simulation study, in which the proposed approach is further evaluated and compared with existing methods from the literature.
\subsection{Academic simulation example}\label{sec:NFIR_sim_example}
To demonstrate the capabilities of the proposed model augmentation approach, we have generated data using the following static nonlinear system:
\begin{equation}\label{eq:sim_data_gen_sys}
    y_k = \theta_0 + \theta_1 u_k + \theta_2u_k^2 + \theta_3 u_k^3 + e_k,
\end{equation}
where $\theta_0=0.01$, $\theta_1=1$, $\theta_2=-0.5$, $\theta_3=0.1$, moreover, $e_k~\sim \mathcal{N}(0, \sigma_\mathrm{e}^2)$. We assume that, based on prior physical knowledge, a to-be-augmented baseline model is available:
\begin{equation}
    \hat{y}_k(u_k) = \hat{\theta}_1 u_k + \hat{\theta}_3 u_k^3,
\end{equation}
where the baseline parameters are $\hat{\theta}_1$, and $\hat{\theta}_2$ with initial values as $\hat{\theta}_1^0 = 0.8$, $\hat{\theta}_2^0 = 0.03$. Similar physics-based baseline models have high relevance in system identification theory \citep{schoukens_identification_2017}. For the learning component, a simple feedforward ANN with 1 hidden layer and 16 neurons is applied, using the \emph{hyperbolic tangent} (tanh) activation. The ANN parameters are initialized with the Xavier method \citep{glorot_understanding_2010}.

Three distinct datasets are generated for training to highlight the importance of data generation, each with $N=1024$ data points. The first approach applies a white noise input signal with $u_k\sim\mathcal{N}(0, 0.3^2)$ for generating 512 samples, then the rest of the data points are acquired by using the first half of the input signal multiplied by -1. This results in a symmetric training data set (denoted by $\mathcal{D}_N^\mathrm{(1)}$), since the even nonlinearities are orthogonal to odd nonlinearities, according to \eqref{eq:phi-delta-discrete-orthogonality}, Condition~\ref{cond:special_PE} is satisfied. The second set $\mathcal{D}_N^{(2)}$ is gathered by applying 1024 input samples directly generated by the previous distribution. This results in a dataset with, most likely, small asymmetries. However, these asymmetries will disappear, and \eqref{eq:phi-delta-discrete-orthogonality} holds for $N\to\infty$. Finally, $\mathcal{D}_N^{(3)}$ is acquired by using an asymmetric distribution of $u_k\sim\mathcal{N}(-0.01, 0.3^2)$, which will not satisfy Condition~\ref{cond:special_PE} when $N\to\infty$. For testing, a data set is constructed with $N_\mathrm{test}=1024$ using the same input distribution as in $\mathcal{D}_N^{(2)}$. For better demonstration of the results, the test data is kept noise-free. To minimize \eqref{eq:cost_fun}, we applied the Adam optimizer for 500 epochs, followed by the L-BFGS method for 1000 iterations, using the identification pipeline proposed in \cite{bemporad_l-bfgs-b_2025}. For benchmarking, we applied the standard additive augmentation, as well as the proposed orthogonal-by-construction method, using 10 different initialization points.

\begin{figure}
    \centering
    \includegraphics[width=0.9\linewidth]{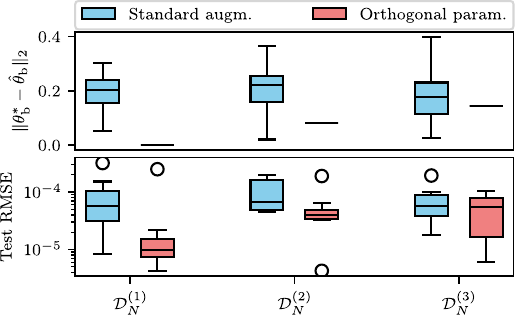}
    \vspace{-6pt}
    \caption{Error of the estimated baseline parameters and RMSE of the simulated model response under $\sigma_\mathrm{e}^2=0$ for 10 Monte Carlo runs.}
    \label{fig:sigma0_errors}
\end{figure}

\begin{figure*}
     \centering
     \begin{subfigure}[b]{0.3\textwidth}
         \centering
         \includegraphics[width=\textwidth]{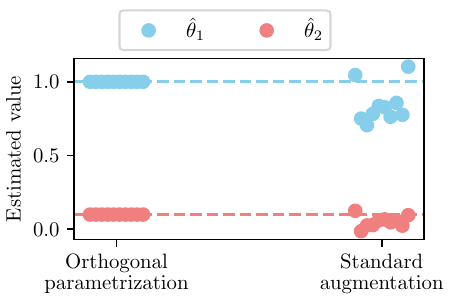}
         \caption{$\mathcal{D}_N^{(1)}$}
         \label{fig:theta_estim_symm}
     \end{subfigure}
     \hfill
     \begin{subfigure}[b]{0.3\textwidth}
         \centering
         \includegraphics[width=\textwidth]{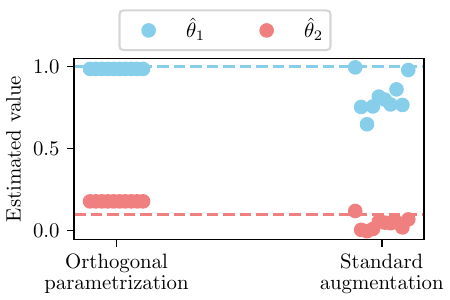}
         \caption{$\mathcal{D}_N^{(2)}$}
         \label{fig:theta_estim_unsymm}
     \end{subfigure}
     \hfill
     \begin{subfigure}[b]{0.3\textwidth}
         \centering
         \includegraphics[width=\textwidth]{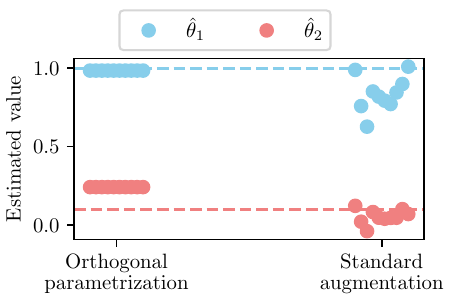}
         \caption{$\mathcal{D}_N^{(3)}$}
         \label{fig:theta_estim_D3}
     \end{subfigure}
     \vspace{-6pt}
        \caption{Estimated baseline parameters with different model structures and training data distributions for $\sigma_\mathrm{e}^2=0$ with 10 Monte Carlo steps. Horizontal dashed lines correspond to the physically true parameter values.}
        \label{fig:theta_estim}
\end{figure*}

\begin{figure*}
     \centering
     \begin{subfigure}[b]{0.3\textwidth}
         \centering
         \includegraphics[width=\textwidth]{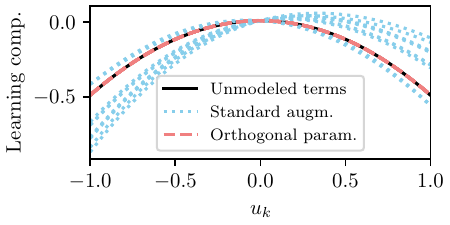}
         \caption{$\mathcal{D}_N^{(1)}$, $\sigma_\mathrm{e}^2=0$}
         \label{fig:learning_comp_plot}
     \end{subfigure}
     \hfill
     \begin{subfigure}[b]{0.3\textwidth}
         \centering
         \includegraphics[width=\textwidth]{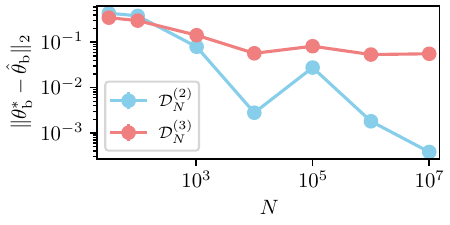}
         \caption{orth. param., $\sigma_\mathrm{e}^2=0$}
         \label{fig:D2_to_infty}
     \end{subfigure}
     \hfill
     \begin{subfigure}[b]{0.3\textwidth}
         \centering
         \includegraphics[width=\textwidth]{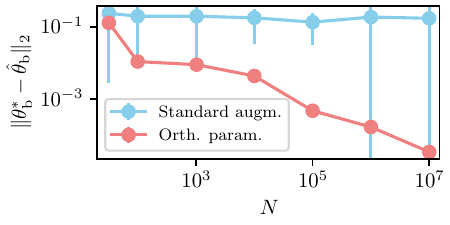}
         \caption{$\mathcal{D}_N^{(1)}$, 30 dB SNR}
         \label{fig:consistency}
     \end{subfigure}
     \vspace{-6pt}
        \caption{Panel (a) shows the learning component outputs compared to the true unmodeled terms for 10 Monte Carlo runs. Panels (b) and (c) show the error of the estimated baseline parameters for different training data lengths. The results are averaged over 10 Monte Carlo runs for each $N$ with the bars representing the $\pm$ two times the standard deviation. These error bars are only visible for the standard augmentation approach, as the orthogonal parametrization brings uniqueness regarding the estimated baseline parameters.}
        \label{fig:infty_cases}
\end{figure*}

First, the data sets were generated using $\sigma_\mathrm{e}^2=0$ to test the methodologies without noise. Based on the 10 different initialization points, the test errors and the estimation errors regarding $\theta_\mathrm{b}$ are shown in Fig.~\ref{fig:sigma0_errors}. Keep in mind that models were estimated on the same data during the Monte Carlo study; only initialization of the model parameters was varied. Both parametrizations with the three different training data sets have generated similar, highly accurate results near the numerical tolerance of the applied optimizer. The main advantage of the orthogonal-by-construction method is highlighted when we investigate the error of the estimated baseline parameters. For $\mathcal{D}_N^{(1)}$, the special persistence of excitation condition, i.e., Condition~\ref{cond:special_PE}, is satisfied, and $\theta_\mathrm{b}$ converges to the physically true values (within numerical error), as shown in Fig.~\ref{fig:theta_estim_symm}. When this condition is not fulfilled, the exact values of $\theta_\mathrm{b}$ can not be recovered, but the uniqueness of the estimated baseline parameter still holds, as shown in Fig.~\ref{fig:theta_estim_unsymm}, and \ref{fig:theta_estim_D3}. It is worth mentioning that using different input sequences sampled from the same distribution would result in slightly different estimation results for $\mathcal{D}_N^{(2)}$ and $\mathcal{D}_N^{(3)}$. For certain ANN initializations, the standard additive structure has yielded reasonably accurate baseline parameters; on average, it provided significantly less accurate $\theta_\mathrm{b}$ estimates, as visible in Fig.~\ref{fig:sigma0_errors}. To show that the ANN with orthogonal projection learns exactly the missing quadratic function in \eqref{eq:sim_data_gen_sys} under Condition~\ref{cond:special_PE}, Fig.~\ref{fig:learning_comp_plot} shows the outputs of the learning component for both the orthogonal parametrization and the standard additive structure (for $\mathcal{D}_N^{(1)}$ and $\sigma_\mathrm{e}^2=0$). It is visible that the proposed method has recovered nearly identically the unmodeled quadratic function for all Monte Carlo runs, whereas the standard approach has generated diverse results depending on the parameter initialization; moreover, most of them exhibit distorted characteristics compared to the true unmodeled terms. We now show that the naive data generation approach used for acquiring $\mathcal{D}_N^{(2)}$ also fulfills Condition~\ref{cond:special_PE} as $N\to\infty$. This is illustrated by repeating the identification task with the orthogonal parametrization for various data lengths. The resulting baseline parameter error values are depicted in Fig.~\ref{fig:D2_to_infty}. As visible, the error converges towards zero for $\mathcal{D}_N^{(2)}$ as $N\to\infty$, on the other hand, it converges towards a fixed number in case of $\mathcal{D}_N^{(3)}$.

\begin{figure}
    \centering
    \includegraphics[width=0.6\linewidth]{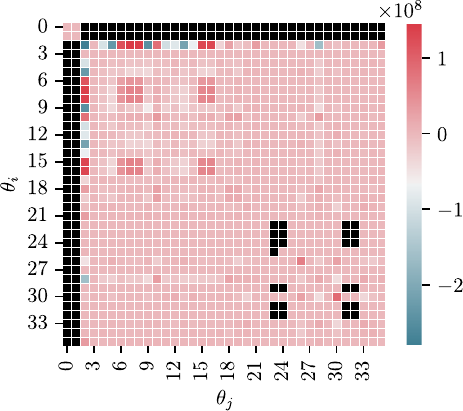}
    \vspace{-6pt}
    \caption{Elements of the asymptotic covariance matrix with (numerically) zero entries shown in black.}
    \label{fig:cov_mx}
\end{figure}

To analyze estimation under noise, the value of $\sigma_\mathrm{e}^2$ was set to reach a \emph{signal-to-noise ratio} (SNR) of 30 dB w.r.t. the measured output. As the importance of satisfying Condition~\ref{cond:special_PE} was demonstrated with the noiseless example, now only $\mathcal{D}_N^{(1)}$ is used. The results are shown in Table~\ref{tab:30dB_errors}. Keep in mind that the test data set does not contain any noise for a better comparison, and similarly as before, only parameter initialization is varied during the Monte Carlo study, the estimation data is kept the same. As expected, a slight increase is visible in the test errors compared to the noiseless scenario, but similarly to before, the orthogonal parametrization and the baseline structure provided nearly identical results considering only model accuracy. On the other hand, the orthogonal-by-construction method resulted in nearly two magnitudes more accurate baseline parameters. For this scenario, the claimed zero covariance is also validated. After training the orthogonal-by-construction structure, $\mathrm{Cov}(\hat{\theta}^N)$ is computed using \eqref{eq:cov_approx}. The values of the asymptotic covariance matrix are depicted in Fig.~\ref{fig:cov_mx} with the elements that are smaller than $10^{-6}$ in absolute value, i.e., numerically zero, shown in black. The first two rows and columns correspond to the baseline parameters, hence the block matrix nature of $\hat{P}^N$ is visible. To showcase the results of Theorem~\ref{thm:error_val}, we repeated the identification task for the noisy case with different training data lengths. As shown in Fig.~\ref{fig:consistency}, the baseline parameter estimation error for the orthogonal-by-construction method clearly converges to zero as $N\to\infty$. In contrast, the standard additive model augmentation approach does not exhibit such convergence behavior, showing the consistency of the proposed parametrization regarding the baseline parameters.

\begin{table}
    \centering
    \caption{Test errors and the error of the estimated baseline parameters with 30 dB SNR.}
    \vspace{-6pt}
    \begin{tabular}{lcc}
    \toprule
        Model & Test RMSE & $\|\theta_\mathrm{b}^\ast - \theta_\mathrm{b}\|_2$\\
    \midrule
        Base additive augm. & $5.86\cdot 10^{-4}$ & 0.1952\\
        Orthogonal param. & $5.35\cdot 10^{-4}$ & 0.0089\\
    \bottomrule
    \end{tabular}
    \label{tab:30dB_errors}
\end{table}

\subsection{Physics-motivated simulation example}
Consider the discretized dynamics of a single-degree-of-freedom \emph{mass-spring-damper} (MSD) system subjected to an external force $u_k$:
\begin{multline}
    y_{k} = \theta_1 y_{k-1} + \theta_2 y_{k-2} + \theta_3 u_{k-1} - \\\theta_4 F_\mathrm{fric}\left(\frac{y_{k-1}-y_{k-2}}{T_\mathrm{s}}\right) + e_k,
\end{multline}
where $y_k$ is the position of the mass, $e_k$ is an i.i.d. white noise process resulting in approximately 30 dB SNR level, $\theta_1$, $\theta_2$, $\theta_3$ and $\theta_4$ are physical parameters that can be expressed using the mass $m$, spring constant $k_\mathrm{s}$, damping coefficient $c_\mathrm{d}$, and sampling time $T_\mathrm{s}$, as
\begin{equation}
    \theta_1 = \frac{2m-c_\mathrm{d}T_\mathrm{s}}{\Gamma},\quad
    \theta_2 = -\frac{m}{\Gamma},\quad
    \theta_3 =\theta_4= \frac{T_\mathrm{s}^2}{\Gamma},
\end{equation}
where $\Gamma = m + c_\mathrm{d}T_\mathrm{s} + k_\mathrm{s}T_\mathrm{s}^2$. Furthermore, the friction force $F_\mathrm{fric}$ is computed according to the Stribeck model, as
\begin{equation}
    F_\mathrm{fric}(v) = F_\mathrm{C}\mathrm{sign}(v) + (F_\mathrm{S} - F_\mathrm{C}) \exp\left(-\frac{\vert v\vert}{v_\mathrm{S}}\right)\mathrm{sign}(v).
\end{equation}
Numerical parameter values corresponding to the data-generating system are $m=1$ kg, $k_\mathrm{s}=1000~\mathrm{N/m}$, $c_\mathrm{d}=100~\mathrm{Ns/m}$, $F_\mathrm{C}=0.8~\mathrm{N}$, $F_\mathrm{S}=1.2~\mathrm{N}$, $v_\mathrm{S}=0.02~\mathrm{m/s}$ with a sampling time of $T_\mathrm{s}=0.01$ s. These values result in $\theta_1\approx 0.48$, $\theta_2\approx -0.48$, $\theta_3=\theta_4\approx 4.8\cdot 10^{-5}$ (all in appropriate SI unit). Similar MSD interconnections have high importance in mechanical engineering practice, as they are used for dynamic modeling a wide range of components, e.g., suspension of a ground vehicle \citep{prabakar_response_2013}. These systems are usually described with linear models, which is convenient for system analysis and control design. Hence, we utilize the following physics-based baseline model:
\begin{equation}
    \hat{y}_{k} = \hat{\theta}_1 \hat{y}_{k-1} + \hat{\theta}_2 \hat{y}_{k-2} + \hat{\theta}_3 \hat{u}_{k-1},
\end{equation}
with $\hat{\theta}_1$, $\hat{\theta}_2$, and $\hat{\theta}_3$ as baseline parameters. We assume that, based on prior knowledge, initial values are available for the parameters as $\hat{m}=1.1$ kg, $\hat{k}_\mathrm{s}=1100~\mathrm{N/m}$, $\hat{c}_\mathrm{d}=90~\mathrm{Ns/m}$. With the sampling time assumed to be known accurately, the resulting initial baseline parameters are $\hat{\theta}_1^0=0.61$, $\hat{\theta}_2^0=-0.52$, $\hat{\theta}_3^0=4.7\cdot 10^{-5}$. We generated the training data set similarly to Sect.~\ref{sec:NFIR_sim_example}, i.e., have generated 5000 samples from a symmetric distribution, as $u_k\sim\mathcal{N}(0, 100^2)$, then the second 5000 samples are generated by multiplying the original sequence with -1. As $y_{k-2}$ and $y_{k-1}$ also affect $y_k$, this experiment design does not fully guarantee orthogonality between the unmodeled dynamics and the baseline model. To achieve that, a more concise data acquisition process would be required. However, the deviation from orthogonality is minimal, so the estimated baseline parameters are expected to remain close to their true values, which is generally sufficient for practical applications. The test set is acquired by sampling 1000 samples of the system response for $u_k\sim\mathcal{U}(-100, 100)$. For better comparison of the results, the test data does not contain any noise. For both data sets, an initial condition of $y_{-1}=y_{0}=0$ is used, then the nonlinear model is forward simulated using the generated input sequences.

\begin{figure}
    \centering
    \includegraphics[width=0.9\linewidth]{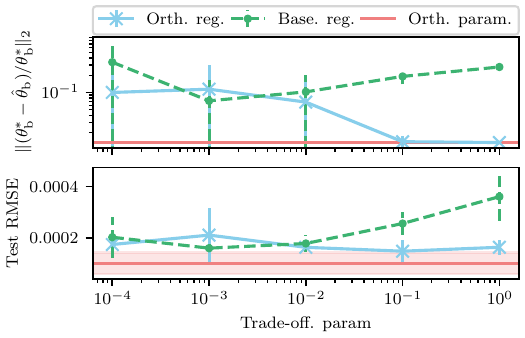}
    \vspace{-6pt}
    \caption{Error of the estimated baseline parameters and RMSE of the simulated model response under different regularization values compared to the proposed orthogonal parametrization. The error bars represent the $\pm$ standard deviation. As the orthogonal-by-construction method does not apply any trade-off parameters, corresponding results are shown with horizontal lines. The shaded area represents the standard deviations.}
    \label{fig:NARX_results}
\end{figure}

We apply the same ANN structure, initialization method, and training environment as in Sect.~\ref{sec:NFIR_sim_example}, but now with 2500 Adam and 500 L-BFGS epochs, respectively. For benchmarking, we apply the standard additive structure, but now complemented with the orthogonal regularization approach (see \cite{kon_physics-guided_2022, gyorok_orthogonal_2025}), which promotes a similar orthogonality as our method. Moreover, for further benchmarking, we apply another regularization-based technique (denoted as "baseline regularization" in the remainder), adapted from \cite{bolderman_feedforward_2022}, which penalizes deviations of the baseline parameters compared to their nominal values. To demonstrate the advantage of our method over regularization-based techniques, we perform a Monte-Carlo study with 5 steps, using various trade-off parameter values (regularization coefficients). As the true baseline model parameters have different magnitudes, now we evaluate the estimation performance by the relative error metric, as $\|(\theta_\mathrm{b}^\ast -\hat{\theta}_\mathrm{b})/\theta_\mathrm{b}^\ast\|_2$. The results are shown in Fig.~\ref{fig:NARX_results}. As shown in the results, the orthogonal regularization–based approach converges toward the proposed method in terms of baseline model parameter estimation error as the trade-off parameter is increased. However, a slight reduction of the overall model performance can be noticed for the regularization-based approach. The baseline regularization method inherently depends on the initial baseline parameter values, since for large values of the trade-off parameter, $\hat{\theta}_\mathrm{b}$ remains close to the initial value $\hat{\theta}_\mathrm{b}^0$. Both approaches demonstrate that, as the regularization coefficient is reduced, the baseline parameters become non-unique and may deviate substantially from the true parameters $\theta_\mathrm{b}^\ast$. At the same time, lower trade-off parameter values can yield improved model performance, an effect that is particularly visible for the baseline regularization method. In practical system identification scenarios, where only test or validation errors are available during hyperparameter tuning, selecting appropriate regularization coefficients is therefore nontrivial and potentially a complex task. This observation highlights a key advantage of the proposed orthogonal-by-construction method, namely that it requires no hyperparameter tuning.

\section{Conclusion}\label{sec:conclusion}\vspace{-6pt}
In this paper, an orthogonal-by-construction parametrization has been introduced for DT IO baseline models. The proposed methodology addresses the challenges of the additive model augmentation structure when the baseline and learning parameters are co-estimated. Detailed theoretical analysis has shown that the orthogonal parametrization can recover the physically true baseline parameter values under the specified identifiability and 
persistence of excitation conditions. Then, these findings have been validated by numerical experiments. Future research may be directed at extending the approach for augmenting baseline models in state-space form, where the general assumption of no available full-state measurement complicates the orthogonal projection of the learning component, thus requiring careful investigation.\vspace{-6pt}

\bibliography{ifacconf}

@ARTICLE{Gevers2009,
author = {M. Gevers and A. S. Bazanella and X. Bombois and L. Mi\v{s}kovi\'{c}},
journal = {IEEE Transactions on Automatic Control},
number = {12},
pages = {2828-2840},
title = {Identification and the information matrix: how to get just sufficiently rich?},
volume = {54},
year = {2009}
}

@article{ljung_perspectives_2010,
	title = {Perspectives on system identification},
	volume = {34},
	number = {1},
	journal = {Annual Reviews in Control},
	author = {Ljung, Lennart},
	year = {2010},
	pages = {1--12}
}

@article{raissi_physics-informed_2019,
	title = {Physics-informed neural networks: {A} deep learning framework for solving forward and inverse problems involving nonlinear partial differential equations},
	volume = {378},
	shorttitle = {Physics-informed neural networks},
	journal = {Journal of Computational Physics},
	author = {Raissi, M. and Perdikaris, P. and Karniadakis, G. E.},
	year = {2019},
	pages = {686--707}
}

@incollection{daw_physics-guided_2022,
	author = "Daw, Arka and Karpatne, Anuj and Watkins, William D. and Read, Jordan S. and Kumar, Vipin",
	title = {Physics-{Guided} {Neural} {Networks} ({PGNN}): {An} {Application} in {Lake} {Temperature} {Modeling}},
	shorttitle = {Physics-{Guided} {Neural} {Networks} ({PGNN})},
	booktitle = {Knowledge {Guided} {Machine} {Learning}},
	publisher = {Chapman and Hall/CRC},
	year = {2022}
}

@book{bohlin_practical_2006,
	edition = {1st},
	series = {Advances in {Industrial} {Control}},
	title = {Practical {Grey}-box {Process} {Identification}},
	publisher = {Springer London},
	author = {Bohlin, Torsten},
	year = {2006}
}

@ARTICLE{schoukens_nonlinear_2019,
  author={Schoukens, Johan and Ljung, Lennart},
  journal={IEEE Control Systems Magazine}, 
  title={Nonlinear System Identification: A User-Oriented Road Map}, 
  year={2019},
  volume={39},
  number={6},
  pages={28-99}
}

@inproceedings{schon_multi-objective_2022,
	title = {Multi-{Objective} {Physics}-{Guided} {Recurrent} {Neural} {Networks} for {Identifying} {Non}-{Autonomous} {Dynamical} {Systems}},
	booktitle = {Proc. of the 14th IFAC Workshop on Adaptive and Learning Control Systems},
	author = {Schön, Oliver and Götte, Ricarda-Samantha and Timmermann, Julia},
	year = {2022},
	pages = {19--24}
}

@inproceedings{djeumou_neural_2022,
  title = 	 {Neural Networks with Physics-Informed Architectures and Constraints for Dynamical Systems Modeling},
  author =       {Djeumou, Franck and Neary, Cyrus and Goubault, Eric and Putot, Sylvie and Topcu, Ufuk},
  booktitle = 	 {Proc. of the 4th Annual Learning for Dynamics and Control Conference},
  pages = 	 {263--277},
  year = 	 {2022}
}

@inproceedings{bolderman_feedforward_2022,
	title = {On feedforward control using physics–guided neural networks: {Training} cost regularization and optimized initialization},
	shorttitle = {On feedforward control using physics–guided neural networks},
	booktitle = {Proc. of the {European} {Control} {Conference}},
	author = {Bolderman, Max and Lazar, Mircea and Butler, Hans},
	year = {2022},
	pages = {1403--1408}
}

@article{bolderman_physics-guided_2024,
	title = {Physics-guided neural networks for feedforward control with input-to-state-stability guarantees},
	volume = {145},
	journal = {Control Engineering Practice},
	author = {Bolderman, Max and Butler, Hans and Koekebakker, Sjirk and Van Horssen, Eelco and Kamidi, Ramidin and Spaan-Burke, Theresa and Strijbosch, Nard and Lazar, Mircea},
	year = {2024},
	pages = {105851}
}

@inproceedings{kon_physics-guided_2022,
	title = {Physics-{Guided} {Neural} {Networks} for {Feedforward} {Control}: {An} {Orthogonal} {Projection}-{Based} {Approach}},
	shorttitle = {Physics-{Guided} {Neural} {Networks} for {Feedforward} {Control}},
	booktitle = {Proc. of the {American} {Control} {Conference}},
	author = {Kon, Johan and Bruijnen, Dennis and Van De Wijdeven, Jeroen and Heertjes, Marcel and Oomen, Tom},
	year = {2022},
    pages = {4377--4382}
}

@inproceedings{gyorok_orthogonal_2025,
	title = {Orthogonal projection-based regularization for efficient model augmentation},
	booktitle = {Proc. of the 7th {Annual} {Learning} for {Dynamics} {\&} {Control} {Conference}},
	author = {Györök, Bendegúz M. and Hoekstra, Jan H. and Kon, Johan and Péni, Tamás and Schoukens, Maarten and Tóth, Roland},
	year = {2025},
	pages = {166--178}
}

@article{hoekstra_learning-based_2025,
    title = {Learning-based model augmentation with {LFR}s},
    journal = {European Journal of Control},
    pages = {101304},
    year = {2025},
    author = {Jan H. Hoekstra and Chris Verhoek and Roland Tóth and Maarten Schoukens}
}

@article{retzler_learning-based_2024,
	title = {Learning-based augmentation of physics-based models: an industrial robot use case},
	volume = {5},
	shorttitle = {Learning-based augmentation of physics-based models},
	journal = {Data-Centric Engineering},
	author = {Retzler, András and Tóth, Roland and Schoukens, Maarten and Beintema, Gerben I. and Weigand, Jonas and Noël, Jean-Philippe and Kollár, Zsolt and Swevers, Jan},
	year = {2024},
	pages = {e12}
}

@article{donati_combining_2025,
	title = {Combining off-white and sparse black models in multi-step physics-based systems identification},
	volume = {179},
	journal = {Automatica},
	author = {Donati, Cesare and Mammarella, Martina and Dabbene, Fabrizio and Novara, Carlo and Lagoa, Constantino M.},
	year = {2025},
	pages = {112409}
}

@inproceedings{glorot_understanding_2010,
	title = {Understanding the difficulty of training deep feedforward neural networks},
	booktitle = {Proc. of the {Thirteenth} {International} {Conference} on {Artificial} {Intelligence} and {Statistics}},
	author = {Glorot, Xavier and Bengio, Yoshua},
	year = {2010},
	pages = {249--256}
}

@article{bemporad_l-bfgs-b_2025,
	title = {An {L}-{BFGS}-{B} {Approach} for {Linear} and {Nonlinear} {System} {Identification} {Under} $\ell_1$ and {Group}-{Lasso} {Regularization}},
	volume = {70},
	number = {7},
    journal = {IEEE Transactions on Automatic Control},
	author = {Bemporad, Alberto},
	year = {2025},
	pages = {4857--4864}
}

@inproceedings{ljung_deep_2020,
	title = {Deep {Learning} and {System} {Identification}},
    booktitle = {Proc. of the 21st {IFAC} {World} {Congress}},
	author = {Ljung, Lennart and Andersson, Carl and Tiels, Koen and Schön, Thomas B.},
	year = {2020},
	pages = {1175--1181}
}

@book{hespanha_linear_2018,
	title = {Linear {Systems} {Theory}: {Second} {Edition}},
	publisher = {Princeton University Press},
	author = {Hespanha, João P.},
	year = {2018}
}

@book{toth_modeling_2010,
	series = {Lecture {Notes} in {Control} and {Information} {Sciences}},
	title = {Modeling and {Identification} of {Linear} {Parameter}-{Varying} {Systems}},
	volume = {403},
	publisher = {Springer},
	author = {Tóth, Roland},
	year = {2010}
}

@article{ljung_convergence_1978,
	title = {Convergence analysis of parametric identification methods},
	volume = {23},
	number = {5},
	journal = {IEEE Transactions on Automatic Control},
	author = {Ljung, L.},
	year = {1978},
	pages = {770--783}
}

@book{ljung_system_1998,
	edition = {2nd},
	title = {System {Identification}: {Theory} for the {User}},
	publisher = {Pearson Education},
	author = {Ljung, L.},
	year = {1998},
}

@inproceedings{huang_leveraging_2024,
    title = {Leveraging {Computational} {Fluid} {Dynamics} in {UAV} {Motion} {Planning}},
    booktitle = {Proc. of the {American} {Control} {Conference}},
    author = {Huang, Yunshen and Greiff, Marcus and Vinod, Abraham and Di Cairano, Stefano},
    year = {2024},
    pages = {375--381}
}

@article{prabakar_response_2013,
    title = {Response of a quarter car model with optimal magnetorheological damper parameters},
    journal = {Journal of Sound and Vibration},
    volume = {332},
    number = {9},
    pages = {2191-2206},
    year = {2013},
    author = {R.S. Prabakar and C. Sujatha and S. Narayanan}
}

@ARTICLE{yuhan_space-filling_2025,
    author={Liu, Yuhan and Kiss, Máté and Tóth, Roland and Schoukens, Maarten},
    journal={IEEE Control Systems Letters}, 
    title={On Space-Filling Input Design for Nonlinear Dynamic Model Learning: A Gaussian Process Approach}, 
    year={2025},
    volume={9},
    number={},
    pages={1868-1873}
}

@article{revay_recurrent_2024,
    title = {Recurrent {Equilibrium} {Networks}: {Flexible} {Dynamic} {Models} {With} {Guaranteed} {Stability} and {Robustness}},
    volume = {69},
    shorttitle = {Recurrent {Equilibrium} {Networks}},
    number = {5},
    journal = {IEEE Transactions on Automatic Control},
    author = {Revay, Max and Wang, Ruigang and Manchester, Ian R.},
    year = {2024},
    pages = {2855--2870}
}

@article{kon_unconstrained_2025,
    title = {Unconstrained {Parametrizations} of {Discrete}-{Time} {Linear} {Input}-{Output} {Models}: {Stability} and {Dissipativity} by {Construction}},
    shorttitle = {Unconstrained {Parametrizations} of {Discrete}-{Time} {Linear} {Input}-{Output} {Models}},
    journal = {IEEE Transactions on Automatic Control},
    author = {Kon, Johan and Tóth, Roland and van de Wijdeven, Jeroen and Heertjes, Marcel and Oomen, Tom},
    year = {2025},
    pages = {1--16}
}

@article{schoukens_identification_2017,
    title = {Identification of block-oriented nonlinear systems starting from linear approximations: A survey},
    journal = {Automatica},
    volume = {85},
    pages = {272-292},
    year = {2017},
    author = {Maarten Schoukens and Koen Tiels}
}
\end{document}